\documentclass[preprint]{aastex}
\usepackage{graphicx}
\usepackage{epstopdf}
\usepackage{bm}
\usepackage{lineno}

\newcommand{\be}{\begin{equation}}
\newcommand{\ee}{\end{equation}}
\newcommand{\nn}{\mbox{} \nonumber \\ \mbox{} }
\newcommand{\ba}{\begin{eqnarray}}
\newcommand{\ea}{\end{eqnarray}}
\newcommand{\om}{\omega}

\renewcommand{\div}{{\rm \,div\,}}

\newcommand{\Bf}{{magnetic field}}

\newcommand{\Bfs}{{magnetic fields}}
\newcommand{\NS}{neutron star}

\newcommand{\Lf} {{Lorentz factor}}

\newcommand\eg{{\it{e.g.}}}

\newcommand\lo{\mathrel{\raise.3ex\hbox{$<$}\mkern-14mu\lower0.6ex\hbox{$\sim$}}}
\newcommand\go{\mathrel{\raise.3ex\hbox{$>$}\mkern-14mu\lower0.6ex\hbox{$\sim$}}}

\begin{document}
\title{Turbulent model of Crab nebula radiation
}

\author{Yonggang Luo$^{1}$,
Maxim Lyutikov$^{1}$,
Tea Temim$^{2}$,  Luca Comisso $^3$\\
$^{1}$ Department of Physics and Astronomy, Purdue University, 525 Northwestern Avenue, West Lafayette, IN, USA 47907\\
$^{2}$ Space Telescope Science Institute, 3700 San Martin Drive, Baltimore, MD 21218, USA\\
$^3$ Department of Astronomy, Columbia University, 550 W 120th St, New York, NY 10027, USA }

\begin{abstract}
We construct a turbulent model of the Crab Nebula's non-thermal emission. The present model resolves a number of  long-standing problems of the Kennel-Coroniti (1984) model:  (i) the sigma problem;  (ii)  the hard spectrum of  radio electrons; (iii) the high peak energy of gamma-ray flares; (iv) and the  spacial  evolution of the infrared (IR) emission. The Nebula contains two populations of injected particles: Component-I  accelerated at the wind termination shock via Fermi-I mechanism, and Component-II accelerated in reconnecting turbulence   in highly magnetized ($\sigma$ $\gg 1$) plasma in the central part of the Crab Nebula. The reconnecting turbulence  Component-II extends from radio to gamma rays: it 
accelerate radio electrons with a hard spectrum,    destroy  the  large scale magnetic flux (and thus resolves the sigma-problem),  and occasionally produce gamma-ray flares (from the  largest scale reconnection events). The  model   reproduces the broad-band spectrum of the Crab Nebula, from low-frequency synchrotron emission in  radio  to inverse-Compton emission at TeV energies,   as well as spatially resolved evolution of the spectral indices in 
 IR and optical bands.
\end{abstract}

\section{Introduction}
\label{Introduction}

\subsection{The  Kennel-Coroniti model:  its success, problems and  resolution}
The Crab  Nebula is the paragon of high energy astrophysical sources - understanding particle acceleration in the Crab has implications for other  sources, like active galactic nuclei and gamma-ray bursts.
Conventionally,  particles in the pulsar wind nebulae (PWNe) are assumed to be accelerated at the pulsar  wind termination shock  \citep{1974MNRAS.167....1R,1984ApJ...283..694K,1984ApJ...283..710K,1996MNRAS.278..525A}. The inferred 
particle spectral index $p = 2.2$,   derived from the non-thermal X-ray synchrotron spectrum, matches the expectations for the  Fermi-I   mechanism
\citep[\eg][]{1987PhR...154....1B}. In addition,
numerical Magnetohydrodynamics (MHD) simulations \citep
{komissarov_04,DelZanna04,2014MNRAS.438..278P,2017SSRv..207..137P}, with the assumed particle acceleration at the termination shock,  reproduce well  the overall X-ray morphology of the PWNe.

However, there are clear drawbacks of the \cite{1984ApJ...283..694K,1984ApJ...283..710K} model. The 
 origin of the radio emitting particles is not addressed. The radio spectrum of Crab PWN has a  spectral index $\alpha$ = 0.3 \citep{1997ApJ...490..291B,2017SSRv..207..175R}, which implies a particle spectral index of $p = 1.6$ for an isotropic distribution of non-thermal electrons.  Such hard radio emission is not consistent  with   the Fermi-I acceleration  mechanism (assumed to be operational at the terminate shock), which typically gives p $>$ 2 \citep[\eg][]{1987PhR...154....1B}.  
  In addition, the lowest observed radio emission from the Crab Nebula, down to 100 MHz, requires {\Lf}s of only $10^2$, well below the typically expected wind \Lf\ of  $\gamma_w \sim 10^4-10^6$  \citep[\eg][]{arons_07,2012SSRv..173..341A}. 

The second major problem in modeling the Crab Nebula's emission, identified by \cite{1974MNRAS.167....1R,1984ApJ...283..694K}, is the  is so-called sigma-problem: models of pulsar magnetospheres  \citep{1977ApJ...217..227F,HardingMuslimov98,Hibschman} predict $\sigma \gg 1$, where sigma is the conventional magnetization parameter \citep[][]{1984ApJ...283..694K}. Supersonic flows with $\sigma \gg 1$  (carrying {\it  large-scale} \Bf)  cannot be accommodated with the non-relativistically expanding nebula. The resolution to the sigma-problem is the destruction of the large-scale magnetic flux, either in the wind \citep{Coroniti90} (but see  \cite{lyubarsky_kirk_01}), or in the  turbulent post-shock flow
 \cite{lb03,2006NJPh....8..119L,2013MNRAS.431L..48P,2017ApJ...847...57Z,2018MNRAS.478.4622T}. We accept the latter interpretation, see Section \ref{sigmaproblem}. 

The third
problem of \cite{1984ApJ...283..694K}  model is related to  Crab's gamma-ray flares \citep{Tavani11,abdo_11,buehler_12}. As discussed by \cite{2010MNRAS.405.1809L} (before the discovery of the flares) and \cite{2012MNRAS.426.1374C} \citep[see also][]{1996ApJ...457..253D}, the peak energy of flares - as high as 400 MeV- violates the synchrotron limit, and is inconsistent with the slow Fermi-I-type acceleration at the shock front. Reconnection in magnetically dominated plasma may accelerate particles at a much faster rate, resolving the problem of the  high-peak energy of flares \citep{zenitani_01,LyutikovUzdensky,lyubarsky_05,2010MNRAS.405.1809L,2012MNRAS.426.1374C,komissarov_12,2008ApJ...682.1436L,hoshino_lyubarsky_12,2014ApJ...782..104C,2017JPlPh..83f6302L,2017JPlPh..83f6301L,2018JPlPh..84b6301L}

The fourth problem of the Kennel-Coroniti model is that it is in
 significant conflict with the observed radial-spectral dependence of the PWNe \citep{2009ApJ...703..662R,2017SSRv..207..175R}.
Models predict a drop in size of the PWN by at least a factor two between radio and X-ray wavelengths,
but  observed PWNe do not show this behavior.

We suggest 
a common resolution to all  the  problems mentioned above (the spectrum of radio electrons, the sigma problem, the high peak energy of gamma-ray flares, and the resolved spectral evolution). We foresee that there are two non-thermally-emitting components in the Nebula: one (Component-I) is accelerated at the termination shock,  and another   (Component-II)  is accelerated in relativistic reconnection events in the bulk of the Nebula, as argued by \cite{2019MNRAS.489.2403L}, see also \cite{2018PhRvL.121y5101C}. 
Component-I abides by the rules of the  \cite{1984ApJ...283..694K,1984ApJ...283..710K} model, with low  magnetization in the equatorial part of the wind.  
Component-II results from 
the highly magnetized  plasma turbulence, which increase the rate of reconnection \cite{doi:10.1063/1.866004},  in the bulk of the nebula and destroys the  magnetic flux  in reconnection events.   The largest reconnection events result in gamma-ray flares \citep{2012MNRAS.426.1374C}.

In \S \ref {sigmaproblem} we discuss the sigma-problem from the point of view of the conservation of large-scale magnetic flux.
In \S \ref {Confinement}  we construct a turbulent model of PWNe. In \S \ref{distribution1} we consider  the evolution of particles in a changing \Bf\ of the Nebula. In \S \ref{reconnectingturbulence}, we discuss the particle acceleration mechanisms in magnetically-dominated reconnecting turbulence.
In \S \ref{Main} we construct the turbulent model of the Crab Nebula radiation. In  \S \ref{indexmap} we construct the corresponding spectral maps in the IR and optical and compare them with observational data.




\section{The sigma-problem -  the problem  of  the magnetic flux}
\label{sigmaproblem}

To clarify the sigma-problem, and to highlight its resolution \citep{lb03,2006NJPh....8..119L}, let us consider a central source (a \NS) that injects  into the Crab Nebula a  highly magnetized, $\sigma \sim 1$, relativistic (supersonic - hence causally disconnected from the source) flow that carries {\it a large-scale toroidal \Bf}. 
If at the injection radius $r_{in}$  ($\sim$ light cylinder) the \Bf\ is $B_{in}$, then the magnetic energy is injected with the rate
\be
\frac{dE_B}{dt} \sim B_{in}^2 r_{in}^2 c
\ee
(for $\sigma \sim 1$, ${dE_B}/{dt}$ is of the order of the spin-down luminosity).
The total injected energy is  then
\be
E_{B} =  B_{in}^2 r_{in}^2 c t
\label{Ein}
\ee
At the same time the central source injects magnetic flux,  integrated over half cross-section of the Nebula, at  a rate
\be
\frac{d\Phi}{dt} \sim B_{in} r_{in} c
\label{Phi}
\ee
(the total injected flux,  integrated over the whole cross-section of the Nebula, is zero, with two opposite contributions of the value (\ref{Phitot}) through two  east-west cross-sections.). 
The total flux,  integrated over half cross-section,  stored in the nebula is
\be
\Phi_{tot} \sim  B_{in} r_{in} c t
\label{Phitot}
\ee

 If the cavity expands with velocity $V_{PWN}$, the \Bf\  and the energy in the bulk are
\ba &&
B \sim \frac{\Phi_{tot}}{ (V_{PWN} t)^2} = \frac{c B_{in} r_{in}   }{ t V_{PWN}^2}
\nn &&
E_{stored} \sim   B^2  (V_{PWN} t)^3 = \frac{ B_{in}^2 r_{in}^2 c^2 t}{V_{PWN}}
\label{Es}
\ea
Comparing (\ref{Ein})  and (\ref{Es}), the injected and the stored energy, it is then required that $ V_{PWN} \sim c$ - only relativistically expanding nebula can accommodate the injected flux. Since PWNe expand non-relativistically  our assumption that a central source injects a highly magnetized relativistic flow leads to an inconsistency - this is the sigma paradox. Only weakly magnetized flows, with magnetic energy flux much smaller than the total wind luminosity  by   $\sigma \sim V_{PWN}/c$, can be matched to the non-relativistically expanding boundary \citep{1984ApJ...283..694K}.

This exercise also suggest  a resolution of the sigma paradox: what is needed is the destruction of the large scale  magnetic flux  (but not necessarily of  the \Bf!).
Consider  a large scale magnetic loop, which has zero total  toroidal flux composed of two opposite contributions in the two east-west cross-sections. If the loop  is broken into small loops, the total flux remains zero, but also now the flux is zero through any east-west cross-sections. Relation (\ref{Phi}) is then not valid any longer
-  there is then no sigma paradox.

Thus, if the \Bf\ is converted into small scale structures, it would behave as  a fluid with some specific equation of state. For example, if a ``fluid'' is composed of magnetic bubbles, then the conservation of flux within a bubble would produce magnetic pressure 
\be
B^2 \propto V_b^{-4/3}
\label{EoS}
\ee
where $V_b$ is the volume of a bubble. This scaling is  reminiscent of the relativistic fluid with adiabatic index of $4/3$. 
\cite{2013MNRAS.431L..48P} indeed demonstrated numerically  that development of current-driven instabilities in the post-termination shock region leads to the resolution of the sigma problem. 

 Given the above arguments, we conclude that instead of smooth flow imagined by  \cite{1984ApJ...283..694K}, the PWNe  {\it must} be highly turbulent. Below we develop a magnetohydrodynamic and radiation model of a PWN, {\it  assuming} it is dominated by turbulence. Previously,  a number of models took into account turbulence and ensuing diffusion on top of the Kennel-Coroniti flow \citep[\eg][]{1972Ap&SS..16...81G,1991ICRC....2..400R,2012ApJ...752...83T,2016MNRAS.460.4135P}. Here we  take an extreme position that magnetohydrodynamic turbulence dominates the flow. This is surely  an extreme assumption: in reality the flow is partially magnetic flux conserving (as demonstrated by large-scale  polarization {structures} that imply toroidal \Bf\ \citep{2008Sci...321.1183D}) and partially turbulent. Yet, as we argue, this extreme 1D  model does reproduce various observational phenomena and resolve the problems of the Kennel-Coroniti model.

\section{Confinement of the turbulent  Crab Nebula PWN by its supernova remnant}
\label{Confinement}

As we argued above, destruction of the magnetic flux is needed to resolve the sigma-problem.  This is achieved via reconnecting turbulence  in the post-shock flow. 
In this Section we  construct a turbulent model of PWNe, whereby the post-shock flow quickly becomes highly turbulent, thus losing the extra requirement of magnetic flux conservation. We consider an extreme case of complete destruction of the magnetic flux. Naturally,  this is an approximation - the real PWN does keep some toroidal magnetic flux, as illustrated by polarized emission from high energy \citep{2008Sci...321.1183D,2016MNRAS.456L..84C,2017NatSR...7.7816C,2018MNRAS.477L..45C} to microwaves \citep{2018A&A...616A..35R,2018arXiv180706207P}, to the radio \citep{1991ApJ...368..231B}.


\subsection{Overall expansion}
\label{overal}

Consider a central source  producing a relativistic supersonic  wind with luminosity $L_w$, confined  within a homologusly expanding stellar envelope. Let us first estimate the overall dynamics of the bubble in the early stages of expansion, when the reverse shock in the ejecta has not yet reached the expanding PWN. 

The stellar envelope  ejected during the supernova explosion expands homologusly, so that its density evolves according to 
\ba &&
\rho = \frac{3}{4 \pi} \frac{M_{ej} } { (V_{ej} t)^3} 
\nn &&
E_{ej} = \frac{3}{10} M_{ej} V_{ej}^2,
\nn &&
v_r = \frac{r}{t},\, r \leq V_{ej} t
\ea
where $ M_{ej}$ is ejecta mass and $V_{ej}$ is the maximal velocity;
a more general scaling of $\rho$ can also be used, $\rho \propto  t^{-3} f(r/t)$, $v_r \propto(r/t) f(r/t)$.

Conventionally \citep[\eg][]{2005ApJ...619..839C} the dynamics of the PWN is treated in what could be called a Sedov approximation, whereby the  internal pressure of the nebular drives supersonic expansion into the supernova ejecta. (Roughly speaking, Sedov approximation is applicable if the size of the termination shock in the pulsar wind is much smaller than the size for the PWN.) In this case the mass, momentum and energy conservation equations are
\ba &&
\partial _t M = 4\pi R^2 \rho \left(V-\frac{R}{t}\right)
\nn &&
M \partial _t V= 4 \pi R^2 \left(p - \rho \left(V-\frac{R}{t}\right)^2\right)
\nn &&
\partial _t (4 \pi p R^3) = L_w - 4 \pi R^2 V p
\nn &&
V= \partial _t R
\ea
($p$ and $\rho$  are  pressure and density internal to the expanding PWN, $L_w$ is wind luminosity, $V$ is overall velocity of expansion.).

The wind luminosity is given by the pulsar spin-down power:
\ba &&
L_w = \frac{I_{NS}  \tau \Omega_0^4}{ 2 ( 1+ t \tau \Omega_0^2)^2}
\label{Lw1}
\nn &&
\tau = 2 \frac{B_{NS}^2 R_{NS}^6 }{ I_{NS} c^3}
\nn &&
\Omega= \frac{\Omega_0}{\sqrt{1+t \tau \Omega_0^2}} = \frac{\Omega_0}{\sqrt{1+t/t_0}}
\nn &&
t_0 = \frac{c^3 I_{NS}}{2 B_{NS}^2 R_{NS}^6 \Omega_0^2}
\ea
where $I_{NS}$ is the moment of inertia of the \NS, $\Omega_0$ is the initial spin, $\Omega$ is the current spin, $B_{NS}=4\times 10^{12}$ G is surface \Bf\ and $R_{NS}=10^6$ cm is radius of the \NS.

As a simplifying assumption in our 1D model, we neglect the evolution of the spin-down power and assume $L_w\sim$ constant. This assumption excludes possible extremely  high initial spins, as suggested by \cite[][so that the population of radio electrons now  is dominated by the very yearly]{1999A&A...346L..49A}. Higher luminosity at earlier times will mildly affect (slightly underestimate)  population of radio emitting electrons.


Assuming constant wind power the corresponding scaling are
\ba &&
R_{PWN} = 0.38 \left(  \frac{L_w V_0^5}{E_{ej}} \right)^{1/5} t^{6/5}= R_{PWN,now}\left(  \frac{t}{t_{now}} \right)^{6/5}
\nn &&
M= 22.4  \left(  \frac{E_{ej}^2 L_w^3} {V_0^{10}}\right)^{1/5} t^{3/5}
\nn && 
p =0.064\left(  \frac{E_{ej}^3 L_w^2} {V_0^3}\right)^{1/5}   t^{-13/5}
\label{RPWN1}
\ea
where $R_{PWN}$ is the radius of the PWN, $M$ is the swept-up mass and $p$ is the pressure.

\subsection{Internal velocity  structure of turbulent PWN flow}
\label{internal}

Let us adopt a limiting case, where instead of smooth flow envisioned by \cite{1984ApJ...283..694K} the requirement of magnetic flux destruction leads to a completely turbulent flow in the nebula. 
The  turbulent \Bf\ behaves as  a fluid, with some specific equation of state, Eq.(\ref{EoS}). The post-shock plasma is relativistically hot, with the sound speed $c_s \sim c/\sqrt{3}$.
The post-shock evolution of the fluid (mixture of relativistic plasmas and turbulent \Bf) will  then quickly reach sub-relativistic velocities and, hence, an incompressible limit.

Consider incompressible flow within a sphere expanding according to (\ref{RPWN1}).
Looking for the  flow velocity of the incompressible fluid in the form
$ v (r,t) = V_{ej}(t) f(x)$ with $x= r/R_{PWN}(t)$, we find 
\be
v = \frac{6}{5}  \frac{R_{PWN,now}^3 t^{13/5}}{r^2 t_{now}^{18/5}}
\label{voft1}
\ee
(this satisfies the condition $\div {\bf v}=0$ and matches to the boundary expansion).
Eq. (\ref{voft1}) gives the velocity of fluid element located at time $t$ at a distance $r$; it is parametrized to the size $R_{PWN,now}$ and age $t_{now}$ of the Crab Nebula now.

The flow should also match the post-termination-shock conditions (\eg, $v_{term. shock} = c/3$ in the purely fluid regime). Clearly this cannot be done in a mathematically  meaningful sense - the system becomes overdetermined. Still, the estimate of the location of the termination-shock,
\be
\frac{r_{term. shock}}{ R_{PWN,now}} \approx \sqrt{ \frac{R_{PWN,now}}{c t_{now}}} \approx 0.1
\ee
is a reasonable estimate of the  relative size of the termination shock with respect to the overall Nebula.  Recall, that one of the effects of the sigma-problem within the model of \cite{1984ApJ...283..694K} was that the size of the termination shock becomes too small for $\sigma \rightarrow 1$. The turbulent model avoids that problem. We consider this as a major advantage of the model.

Consider  next a shell ejected  at time $t_{ej}$ from  the termination shock of radius $R_{ej}$.  Integrating equation of motion (\ref{voft1}) with $v= dr/dt$, the location of the shell at time $t$ is
\ba &&
\frac{r_{shell}}{R_{PWN,now}} =
\left(\left(  \frac{R_{ej}}{R_{PWN,now}} \right)^{3} + \left(  \frac{t}{t_{now}} \right)^{18/5} - \left(  \frac{t_{ej}}{t_{now}} \right) ^{18/5} \right)^{1/3}
\rightarrow 
\nn &&
\left(\left(  \frac{R_{ej}}{R_{PWN,now}} \right)^{3} + 1 - \left(  \frac{t_{ej}}{t_{now}} \right) ^{18/5} \right)^{1/3}
\label{rshell11}
\ea
(A a check, for $t_{ej}=0$  and $R_{ej}=0$ Eq. (\ref{rshell11}) reproduces (\ref{RPWN1})). The last equality in (\ref{rshell11}) refers to the present time, $t=t_{now}$.)

A shell located at $ r_{shell,now }$ at present time  has been ejected at time
 \be
\frac{ t_{ej} }{t_{now}} = \left(1+\left(  \frac{R_{ej}}{R_{PWN,now}} \right)^{3}  - \left(  \frac{r_{shell,now}}{R_{PWN,now}} \right) ^{3} \right)^{5/18}
\label{teject}
\ee

\subsection{Magnetic field within the shell}
\label{internal1}

At each moment the amount of the energy injected by the pulsar should balance nebula pressure, given by the sum of magnetic and kinetic pressures $p_k$. 
(Plasma within the Nebula is relativistically hot, hence we can neglect the  energy of the bulk motion which is smaller by a factor $(v/c)^2$ than the combined enthalpy.)

Using (\ref{RPWN1}) with total pressure given by the sum of kinetic and magnetic pressure,  
\be
p_{tot} = \frac{B^2}{8\pi} +p_k= \frac{B^2}{8\pi} (1+\beta)
\ee
where $\beta$ is the plasma beta parameter, the \Bf\ within a nebula at time $t$ is then
\ba &&
B(t)= B_{now}\left(  \frac{t}{t_{now}} \right)^{-13/10}
\nn &&
B_{now}= 16.4 \frac{E_{ej}^{3/10} L_w^{1/5}}{V_0^{3/2} \sqrt{1+\beta} }t_{now}^{-13/10}=
\frac{\sqrt{6 L_w t_{now}}}{{R_{PWN,now}^{3/2}}} \approx 
6 \times 10^{-4} \, {\rm G}
\label{Boft}
\ea
where the last estimate assumes ejecta energy $E_{ej} =10^{51}$ ergs, maximum velocity $V_0 = 7500$ km s$^{-1}$ and $\beta =10^2$. 

Given the nature of the order-of-magnitude estimates, the above values is very close to the estimates of the \Bf\ in the Nebula \citep[\eg][]{2017SSRv..207..175R}. We consider this as another major advantage of the model.

\section{Particle distribution within the nebula}
\label{distribution1}

Above, we constructed a fluid-like  turbulent model of PWN, composed of shells of material injected at different times. Magnetic field in each shells evolves with time according to (\ref{Boft}). 
In this Section we calculate the radiation signatures of  such turbulent PWN.
 In \S \ref{distr} we  consider the evolution of the particle distribution within each injected shell, taking into account  radiative losses (there are no adiabatic losses in the incompressible approximation).
 
In subsection \ref{distr}, we  find the Green's function for particles injected at some moment and an experiencing radiative decay in an evolving \Bf. The Green's function, multiplied by the injection rate,  gives the  particle distribution function within each shell. Next,  in subsection \ref{overdis}, we integrate the  Green's function over the injection time to find  the total particle distribution within the Nebula. 
 
\subsection{Evolution of the particle distribution in a changing magnetic field}
\label{distr}

We assume that particles are injected into the inner regions of the PWN with some given distribution and seek to 
 find the particle distribution within each injected shell, taking into account radiative losses and a changing \Bf\ within each shell.   We need to
 solve the  Boltzmann's  (Liouville's) equation for the Green's function $G$
\begin{equation}
\frac{\partial G}{\partial t}= \frac{\partial (\dot{\gamma}G)}{\partial {\gamma}}+f_{inj}\delta(t-t_{inj})
\label{Liouville}
\end{equation}
for
an injected spectrum with a power-law particle distribution
\be
f_{inj} \propto \gamma_{inj}^{-p},\, \gamma>  \gamma_{inj, min},
\ee
where $t_{inj}$ is the moment of injection and  $\gamma_{inj, min}$ is a minimum injection \Lf.

Consider first the evolution of the \Lf\ of the particles experiencing radiative losses in an evolving \Bf,
\ba && 
\dot{\gamma}  = - \frac{4}{9}  \frac{e^2}{m_e c^3} \gamma^2 \om_B^2
\nn && 
\om_B=\frac{e B}{m_e c}
\nn &&
B = {B_0} \left(\frac{t}{t_0}\right)^{-\delta}
\label{dgamma}
\ea
with $\delta > 1/2 $. (In our case, $\delta=13/10$, see Eq.(\ref{Boft}).) For definiteness we can set $t_0=t_{now}$, so that $t<t_0$. 

Introducing 
\ba &&
\tau_c=  \frac{9}{4} \frac{m_e^3 c^5}{e^4 {B_{now}}^2}
\nn &&
\gamma_M = \frac{ \tau_c}{t_{now}},
\label{gammaM}
\ea
Eq. (\ref{dgamma}) can be written as
\be
\dot{\gamma}  = - \left(\frac{t_{now} }{t} \right)^{2 \delta} \frac{\gamma^2}{t_{now} \gamma_M}
\ee
If at time $t_{inj}$ a particle was injected with \Lf\ $\gamma_{inj}$, then the \Lf\ evolves according to
\ba &&
\frac{\gamma}{\gamma_{inj}}= \left(1 +\frac{1}{2 \delta-1}  \left( \left( \frac{t_{now}}{t_{inj}} \right)^{2\delta-1}- \left( \frac{t_{now}}{t}\right) ^{2\delta-1}\right) \frac{\gamma_{inj}}{\gamma_M}\right)^{-1}
\nn &&
\frac{\gamma_{inj}}{\gamma}= \left(1 -\frac{1}{2 \delta-1}  \left( \left( \frac{t_{now}}{t_{inj}} \right)^{2\delta-1}- \left( \frac{t_{now}}{t}\right) ^{2\delta-1}\right) \frac{\gamma}{\gamma_M}\right)^{-1}
\label{gammaoft}
\ea

For a given time $t$ the \Lf\ must be smaller than
\be
\gamma_{max}(t) = (2 \delta-1)   \left( \left( \frac{t_{now}}{t_{inj}} \right)^{2\delta-1}- \left( \frac{t_{now}}{t}\right) ^{2\delta-1}\right)^{-1} \gamma_M
\ee
and larger than 
\be
\gamma_{min} (t) =   \left(1+ \frac{1}{2\delta-1} \left( \left( \frac{t_{now}}{t_{inj}} \right)^{2\delta-1}- \left( \frac{t_{now}}{t}\right) ^{2\delta-1}\right)  \frac{\gamma_{inj, min}}{ \gamma_M}\right)^{-1}  \gamma_{inj, min}
\ee

\begin{figure}[t]
  \centering
  \includegraphics[width=0.8\textwidth]{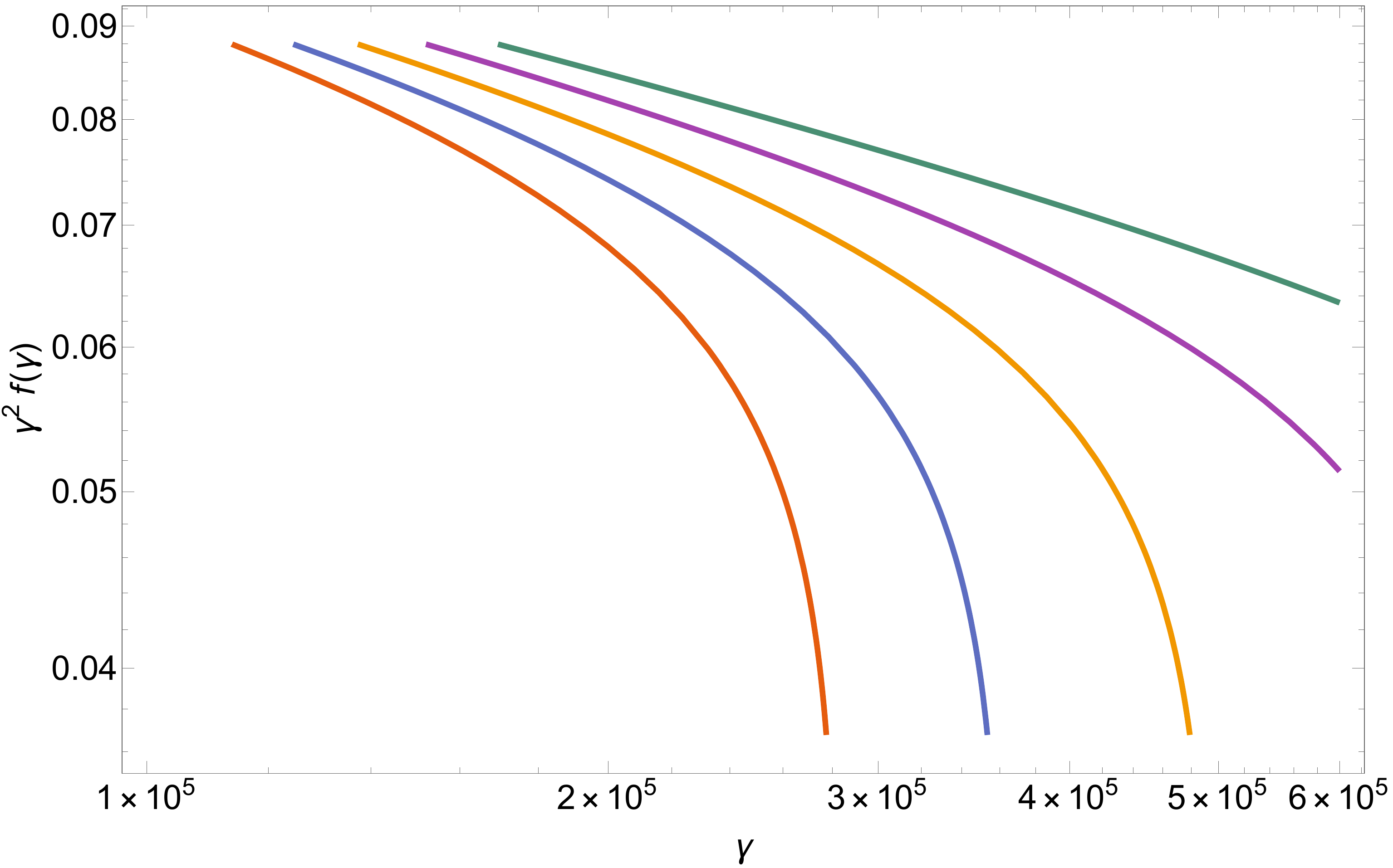}
  \caption{Evolution of the distribution function within one shell. Each line has injection time $t_{inj}$ as $t_{now}/t_{inj}$ =1.1, 1.2, 1.3, 1.4, 1.5 (from green to red) with the same minimum injection Lorentz factor $\gamma_{inj,min}$ and normalization factor. As the particle distribution function evolves with time, particles are cooled due to synchrotron emission and shifted to lower energy. Here power-law index p = 2.2 and the minimum injection Lorentz factor $\gamma_{inj,min} = 1.9 \times 10^5$.}
  \label{distribution}
 \end{figure}

Thus, at any time $t$, the distribution function for particles injected at  $t_{inj}$  is given by 
\be
G (t,t_{inj}) \propto  \gamma^{-p} \left(1-\frac{1}{2\delta-1} \left( \left( \frac{t_{now}}{t_{inj}} \right)^{2\delta-1}- \left( \frac{t_{now}}{t}\right) ^{2\delta-1}\right)  \frac{\gamma}{ \gamma_M} \right) ^{p-2}
\Theta\left(\gamma - \gamma_{min} (t)\right) \Theta\left(\gamma_{max} (t)-\gamma\right) ,
\label{foft}
\ee
see Fig. \ref{distribution}.
Eq. (\ref{foft}) gives the Green's function for the evolution of the particle distribution function.

There is a special injection  time $t_{inj, full}$ so that now, at $t=t_{now}$, for $t_{inj}< t_{inj, full}$ the highest possible \Lf\ becomes smaller that the minimal injection \Lf\ $\gamma_{inj, min}$: in this regime all the particles enter the fast cooling regime:
\be
\frac{t_{inj, full}}{t_{now}} = \left( 1+ (2 \delta -1) \frac{\gamma _M}{\gamma_{inj, min}}\right) ^{-1/(2\delta -1)} \rightarrow
\left(1+ \frac{\gamma _M}{\gamma_{inj, min} } \right)^{-1}
\ee
If $t_{inj} < t_{inj, full}$, then all the particles within a shell cool below $\gamma_{inj, min}$.
Since   ${\gamma _M} \leq {\gamma_{inj, min} }$ most of the particles that have been accelerated above $\gamma_{inj, min}$ over the lifetime of the Nebula had time to cool down below $\gamma_{inj, min}$.

The ratio $\gamma_{max}/   \gamma_{min} $ is
\be
\frac{\gamma_{max}}{  \gamma_{min} } =
1+ (2 \delta-1)  \left( \left( \frac{t_{now}}{t_{inj}} \right)^{2\delta-1}- \left( \frac{t_{now}}{t}\right) ^{2\delta-1}\right)^{-1}  \frac{ \gamma_M}{\gamma_{inj, min}}
\ee
For earlier  $t_{inj} \rightarrow 0$ the ratio $\gamma_{max}/   \gamma_{min} \rightarrow 1$. Thus, with time all the particles injected at some $t_{inj}$ occupy a narrower and narrower range of $d\gamma$ -  there is an effective pile-up in the distribution.

\subsection{The overall particle distribution in the Crab Nebula}
\label{overdis}

\begin{figure}[t]
  \centering
  \includegraphics[width=0.8\textwidth]{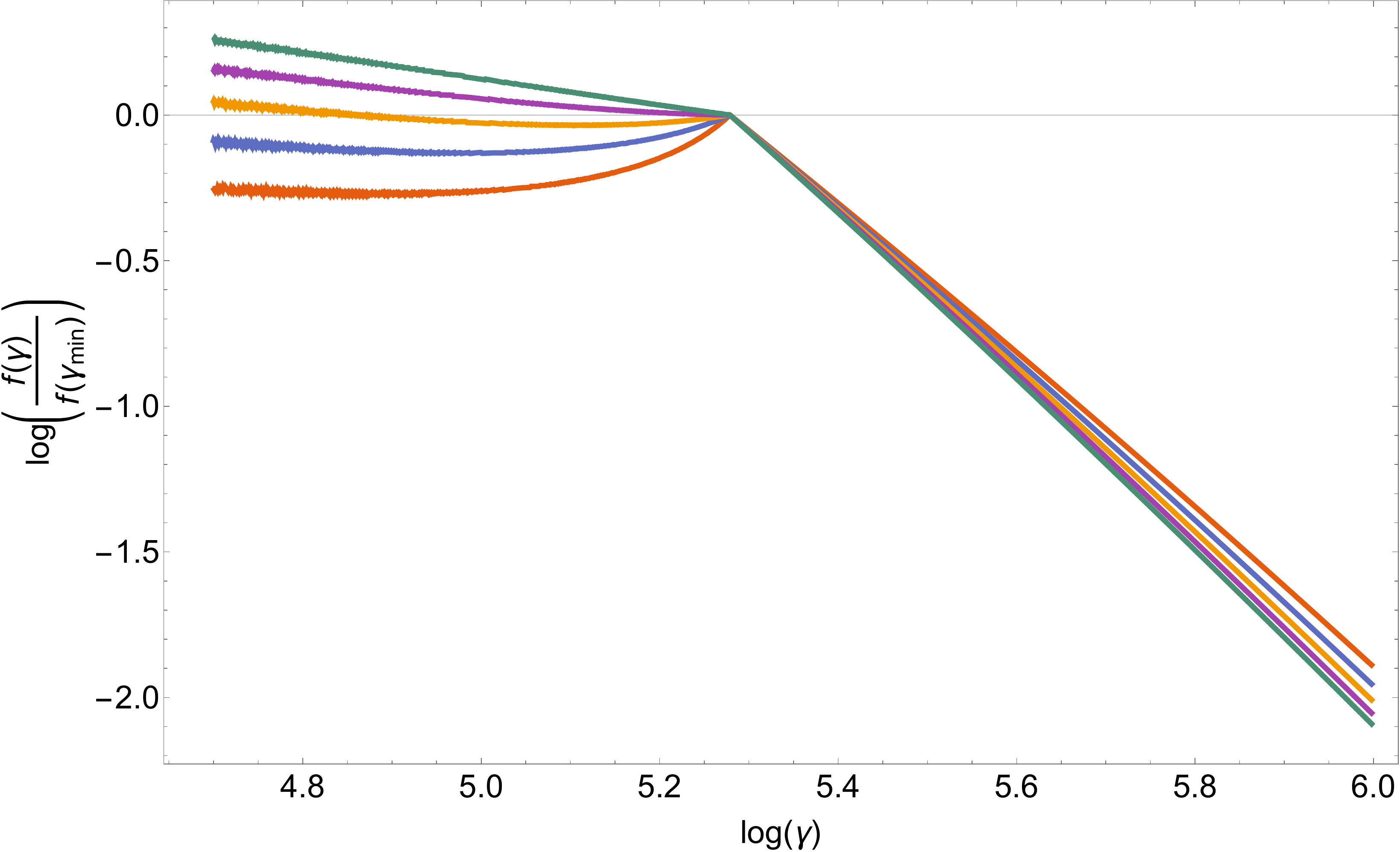}
  \caption{Total particle distribution function within the Nebular for different present-time  \Bfs: $2.0 \times 10^{-4}$G (red),  $2.5 \times 10^{-4}$G (blue),  $3.0 \times 10^{-4}$G (orange), $3.5 \times 10^{-4}$G (purple) and $4.0 \times 10^{-4}$G (green) at $t_{now}$. We keep injecting power-law particle distribution from injection time $t_{inj} = 0.1 \times  t_{now}$ with p = 2.2 and same minimum Lorentz factor $\gamma_{inj,min} = 1.9 \times 10^5$ and let all particles evolve with time. All curves are normalized to unity at the injection break.}
  \label{totaldistribution}
\end{figure}

Eq. (\ref{foft}) describes the evolution of the distribution function for the particles injected at time $t_{inj}$. To find  the total distribution function in the Nebula, the Green's function (\ref{foft}) should be integrated over injection times $t_{inj} \leq t_{now}$.   Results of numerical integration are plotted in   Fig. \ref{totaldistribution} (constant injection parameters are assumed). 

In Fig. \ref{totaldistribution}, there is one injection break at $\gamma_{inj,min}$ for all curves since they all have same minimum injection Lorentz factor. For large magnetic fields (e.g. purple and green curves), particles cool quickly, so that the distribution increases below the injection break towards smaller Lorentz factors and has relatively higher number of particles at lower energy. For small magnetic fields (e.g. red, blue and orange curves), the distribution is nearly constant and has a relatively lower number of particles at lower energies, which are the particles cooled quickly early-on when the magnetic field was strong.

\section{Acceleration in relativistic reconnecting turbulence}
\label{reconnectingturbulence}

In addition to providing a satisfactory solution of the sigma-problem, magnetized turbulence in the bulk of the Crab Nebula is expected to accelerate particles far out of thermal equilibrium. \footnote{To be clear, our model is different from   ''turbulent reconnection'' of \cite{1999ApJ...517..700L}, in that case  ''turbulent reconnection'' is understood as turbulence inside a  reconnecting current sheet.  In contrast, what we envision can be described as turbulence with reconnection occurring in various current sheets inside the turbulence itself.}  Particle acceleration can occur due to a combination of turbulence fluctuations and magnetic reconnection events that are self-consistently produced by the turbulent motions in the plasma. Indeed, in magnetized turbulence, contrary to hydrodynamic turbulence, the presence of the magnetic field gives rise to turbulence eddies that becomes progressively more anisotropic towards small scales within the inertial range, producing current-sheet-like structures that are prone to magnetic reconnection \citep{Carbone_1990,Mallet_2017,Loureiro2017,Comisso_2018} due to the plasmoid instability that kicks in while current sheets are forming \citep{comisso_16,uzdensky_16,Comisso_ApJ2017}.

Recent first-principle kinetic simulations \citep{ComissoSironi18,ComissoSironi19} have shown that in a strongly magnetized plasma ($\sigma \gg 1$), such as the case for the central part of the Crab Nebula, the interplay between turbulence fluctuations and magnetic reconnection leads to the generation of a large fraction of nonthermal particles. The resulting particle energy distribution had been shown to display a power-law energy tail $dn/d\gamma \propto \gamma^{-p}$ that extends well beyond the Lorentz factor 
\begin{equation} \label{gamma_st}
\gamma \sim \left( {1 + {{\sigma}}} \right) \gamma_{0} \, ,
\end{equation}
which takes into account the fact that most of the magnetic energy is converted to particle energy by the time the particle energy spectrum has saturated \citep{ComissoSironi18,ComissoSironi19}. The slope $p$ of the particle energy spectrum was found to depend on the plasma magnetization $\sigma$ and the amplitude of the turbulence fluctuations $\delta B_{{\rm{rms}}}$ with respect to the mean magnetic field $B_0$. In particular, the power-law slope $p$ is harder for larger magnetizations and stronger turbulence fluctuations \citep{zhdankin_17,ComissoSironi18,ComissoSironi19}. For $\sigma \gg 1$ and large turbulent fluctuations ($\delta B_{{\rm{rms}}}^2/ B_0^2 \sim 6$ in some regions of the Crab Nebula, as discussed in \citealt{2019MNRAS.489.2403L}), the power-law slope was found to be $p < 2$ \citep{ComissoSironi18,ComissoSironi19}, but generally not as hard as the slope generated by reconnection alone with the same parameters, which can approach $p\rightarrow1$ for $\sigma \gg 1$ \citep{zenitani_01,ss_14,guo_14,werner_16,2017JPlPh..83f6301L,2017JPlPh..83f6302L}. Therefore, it is conceivable to assume a space-averaged spectrum with a slope $p \sim 1.6$, as can be inferred from the radio spectrum of the Crab Nebula.

More specifically, \citet{ComissoSironi18,ComissoSironi19} have shown that plasmoid-mediated reconnection controls the initial acceleration of particles from the thermal bath at $\gamma_0$ up to the Lorentz factor $\gamma_{0} \left( {1 + {{\sigma}}} \right) $. In our model, $\gamma_0$ corresponds to the wind Lorentz factor in the absence of dissipation. Then, some particles are further accelerated to much higher energies by stochastic interactions with turbulent fluctuations, with the most energetic particles reaching 
\begin{equation} \label{gamma_max}
\gamma_{\rm{max}} \sim \frac{e B_{{\rm{rms}}} {\ell}}{m_e c^2}  \, ,
\end{equation}
where $\ell$ indicates the size of the largest turbulent eddies and $B_{{\rm{rms}}}$ is the space-averaged root-mean-square value of the magnetic field. This two-stage acceleration process is characterized by a combination of systematic (Fermi-I) and stochastic (Fermi-II) particle acceleration mechanisms.

At small scales, the non-ideal reconnection electric fields, whose magnitude is $|E_{\parallel}|\simeq \beta_R \delta B_{\rm rms}$, accelerate particles according to 
\begin{equation} \label{eqRECONN}
\frac{d \langle \gamma \rangle}{dt} =  \frac{e}{m_e c} \beta_R {\delta B_{\rm rms}} \, ,
\end{equation} 
where $\beta_R$ is the average reconnection rate, which is an $O(0.1)$ quantity for relativistic collisionless plasmas \citep{zenitani_hesse_09,bessho_12,  Cerutti2012ApJL,guo_14,kagan_15,liu_15,sironi_16,ComiBhatta2016,werner_17,2017JPlPh..83f6301L}. The fast reconnection rate $\beta_R \sim 0.1$ guarantees that magnetic reconnection can process large volumes of plasma in few outer-scale eddy turnover times, in addition to enabling particle acceleration on a fast timescale $t_{acc} \sim \beta_R^{-1} \rho_L/c$, where $\rho_L$ is the particle Larmor radius.

After the initial acceleration due to plasmoid-mediated reconnection, particles are further accelerated by stochastic scattering off turbulent fluctuations in the inertial range of the turbulent energy cascade. The mean particle energy gain due to stochastic acceleration is related to the diffusion coefficient in energy space as 
\begin{equation} \label{}
\frac{d \langle \gamma \rangle}{dt} = \frac{1}{\gamma^2} \frac{\partial }{\partial \gamma} \left( {\gamma^2 D_{\gamma \gamma}} \right) \, , 
\end{equation} 
with an energy diffusion coefficient $D_{\gamma \gamma}$ that depends on the instantaneous plasma magnetization and the particle Lorentz factor as \citep{ComissoSironi19}
\begin{equation} \label{Dgamma_numerical}
D_{\gamma \gamma}  \sim  0.1 \sigma  \, \left( {\frac{c}{l}} \right) \, \gamma^2  \, ,
\end{equation}
akin to the original Fermi-II mechanism \citep[e.g.][]{blandford_eichler_87,Lemoine19}. Note that the timescale $t_{acc}$ of the stochastic acceleration process is comparable to that of fast plasmoid-mediated reconnection in the strong turbulence scenario considered here. Indeed, the stochastic acceleration timescale is $t_{acc} \sim \gamma^2/D_{\gamma \gamma} \sim 10 \, {\ell}/{\sigma c} $, with $\sigma$ being the instantaneous magnetization. The instantaneous magnetization decreases rapidly in time as a result of magnetic dissipation and reaches $\sigma \sim 1$ in few outer-scale eddy turnover times. Then $t_{acc} \sim 10 \, {\ell}/{c}$ as it would be in the case of fast reconnection ($\beta_R \sim 0.1$) driving particles up to the highest energies allowed by the system size (i.e., with particle Larmor radius $\rho_L \sim \ell$).

Finally, we also expect that at the largest scales, magnetic reconfigurations can generate large scale current sheets whose statistic is not well described as a self-similar sequence controlled by turbulent motions. In this case, the reconnection of the large scale magnetic field might be responsible for particle acceleration up to the maximum available potential. Particle acceleration at these large-scale current sheets can extend up to the synchrotron burn-off limit of 100 MeV and beyond, thus powering the Crab Nebula gamma-ray flares \citep{2017JPlPh..83f6301L,2018JPlPh..84b6301L}. Therefore, in this model of the Crab Nebula radiation, magnetized turbulence with reconnecting current sheets can accelerate both the radio electrons and also produce the Crab gamma-ray flares.

\section{The turbulent model of the Crab Nebula radiation}
\label{Main}

\subsection{Model parameters} 
\label{param}

Above, in Sections \ref{Confinement} and  \ref{distribution1}, we described the one-dimensional  spacial and temporal evolution of the  flow  and of  the distribution function of the accelerated particles as  functions of injection time and the \Bf\ at present time in the Nebula. In this Section, we calculate the resulting broadband spectrum: the synchrotron component and the inverse-Compton component of the non-thermal  synchrotron emission, thermal dust emission, CMB, and starlight photons.


Following \cite{2019MNRAS.489.2403L}
 we assume that there are two acceleration mechanisms in the Crab Nebula: those from  the terminate shock  (Component-I) and the reconnecting turbulence acceleration mechanism  (Component-II).
\citep[The possibility of having
two acceleration mechanisms in PWNe has been suggested previously by][]{1984ApJ...283..710K,1996MNRAS.278..525A,2002A&A...386.1044B,2010A&A...523A...2M,2014PhPl...21e6501C,2013MNRAS.433.3325S,2014MNRAS.438..278P,2014MNRAS.438.1518O,2015MNRAS.449.3149O}.

The Component-I obeys the usual  acceleration conditions of Fermi-I acceleration at the equatorial part of the pulsar wind, the properties of the Component-II are discussed in \S \ref{reconnectingturbulence}.
Both components are accelerated within the inner regions of the Nebula; though Component-II has more extended acceleration cites, see Fig 4 in \cite{2019MNRAS.489.2403L}. Here we neglect the difference in the sizes of the acceleration regions. With time, both components expand hydrodynamically and experience radiative cooling. Component-I is in the fast cooling regime, meaning that particles with the minimal injected energy cool efficiently on the timescale of the PWN. Component-II is from magnetic reconnecting turbulence and is in the slow cooling regime, so that particles with minimal injected energy do not cool.

\begin{figure}[t]
  \centering
  \includegraphics[width=0.8\textwidth]{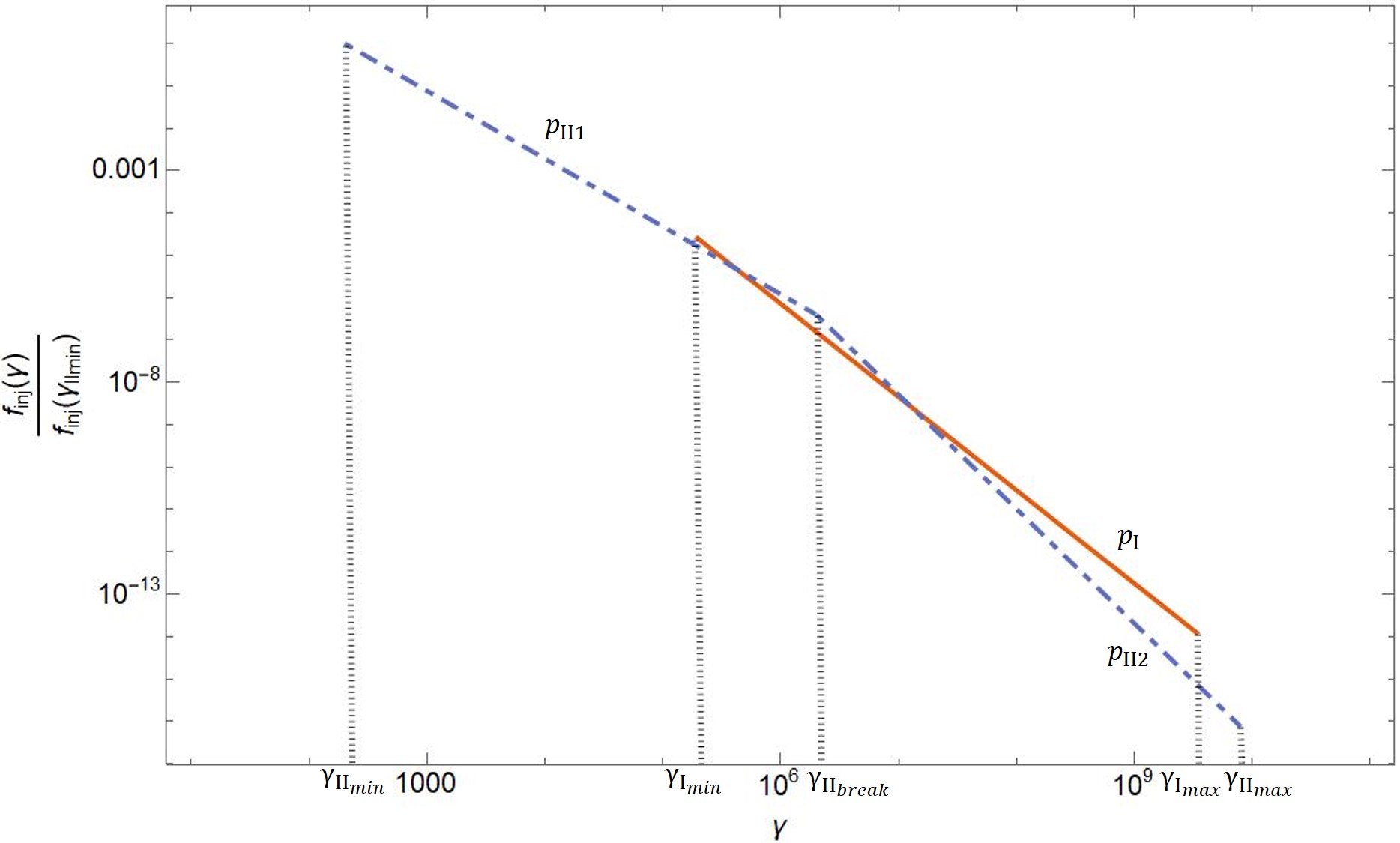}
  \caption{Illustration of parameters in Component-I and Component-II. Component-I is represented by red solid curve and Component-II is represented by blue dashed curve. All parameters values are taken from Table \ref{values}, and we normalized the curve of Component-II to unity at its corresponding minimum injection Lorentz factor $\gamma_{II_{min}}$.}
  \label{injection}
 \end{figure}

We assume that two populations of accelerated particle are injected in the inner region of the Nebula, Fig. \ref{injection}.
The Component-I's injected electron distribution has power-law index $p_{I}$, minimum and maximum injection Lorentz factors $\gamma_{I_{min}}$ and $\gamma_{I_{max}}$. {The values of $p_{I}$  is restricted by the observed spectral power-law indices in the X-ray range, and the value of $\gamma_{I_{min}}$ is restricted by the observed peak and spectral power-law indices in the IR range}. The maximum injection $\gamma_{I_{max}}$ is limited both by the observed break, and the theoretical limit of synchrotron acceleration/burn-off, around 100 MeV \citep[\eg][]{2010MNRAS.405.1809L}.

For Component-II, the injected electron distribution has  a  broken power-law spectrum with indices $p_{II1}$ and $p_{II2}$, minimum and maximum injection Lorentz factors $\gamma_{II_{min}}$ and $\gamma_{II_{max}}$, and break injection $\gamma_{II_{break}}$;  $p_{II1}$ is the power-law index below the injection break $\gamma_{II_{break}}$, $p_{II2}$ is the power-law index above $\gamma_{II_{break}}$. The minimum injection $\gamma_{II_{min}}$ is not restricted: it should be sufficiently low, $\sim$ few hundreds at most, to have the radio spectrum extending down to below $\sim$ 100 MHz. The maximum injection $\gamma_{II_{max}}$ is similarly limited by the acceleration/burn-off. We illustrate these parameters in Fig. \ref{injection}. The spectrum below the break is determined by the observed radio spectrum. The break (approximately in the IR) is required for the Component-II not to overshoot Component-I in the soft X-rays. (In the hard X-rays and gamma-rays the two components contribute similarly).

In our calculation, we fix $p_{I}=2.2$ (this is derived from the X-ray spectrum of the Crab Nebula wisps), $p_{II1}=1.6$ (which is derived from the radio spectral index $\alpha_r=0.3$), and $\gamma_{II_{min}}=200$ (corresponding to synchrotron frequency below few tens of MHz).
There are several fit parameters:  \Bf\ at present time
 $B_{now}$, $\gamma_{I_{min}}$, $\gamma_{II_{break}}$, $\gamma_{I_{max}}$, $p_{II2}$, $\gamma_{II_{max}}$, the relative normalization factor of Component-I and Component-II and the overall normalization factors for each component. 
We explored these parameters and tried to fit the observational data of the IR index map, optical index map, and the broad-band spectrum.

In the following sections, we first calculate the synchrotron  spectrum in \S \ref{Step}, and then the  corresponding IC signal in \S \ref{ic}. The overall spectrum and its evolution is calculated in \S \ref{overall}, and the spatial evolution of spectral indices in the optical and radio in \S \ref{indexmap}.


\subsection{The fitting procedure}
\label{fitting} 

Fitting the broad-band spectrum involving synchrotron and  SSC components as well as other contribution for soft photons (\eg, dust, starlight and CMB) involves numerous parameters and data measurements over a huge range of energies. This  is a complicated task, that cannot be achieved in one-go. Next we describe a novel procedures we developed to tackle this problem. It is somewhat akin to a boot-strap method, where numerous parameters are improved step-wise, trying to achieve the best fit.  

Both Components produce synchrotron emission, and, in addition, there are IC emission on the synchrotron photons (SSC), thermal dust emission, external star light and CMB. A wide range of particles and photons energies requires that KN effects be taken into account for the IC component.  Next, we describe a  novel  procedure to self-consistently fit the synchrotron and IC processes (see \S \ref{ic}) due to two particle distributions.

\subsubsection{The synchrotron component}
\label{Step}

We use the exact expression for local single particle spectral emissivity \citep{rybicki_lightman_79}
\ba && 
P(\omega, r,t) =\frac{\sqrt{3}}{2\pi}\frac{B e^3}{m c^2} F (\frac{\omega}{\omega _c})
\nn &&
\omega _c=\frac{3   }{2 }\gamma ^2 \frac{e B}{ m_e c}
\nn &&
F (x)\equiv x \int_x^{\infty } K_{\frac{5}{3}} (\xi)  \, d\xi
\ea
where $K_{\frac{5}{3}} (\xi)$ is a Bessel function of the second kind.

\begin{figure}[t]
\includegraphics[width=.8\columnwidth]{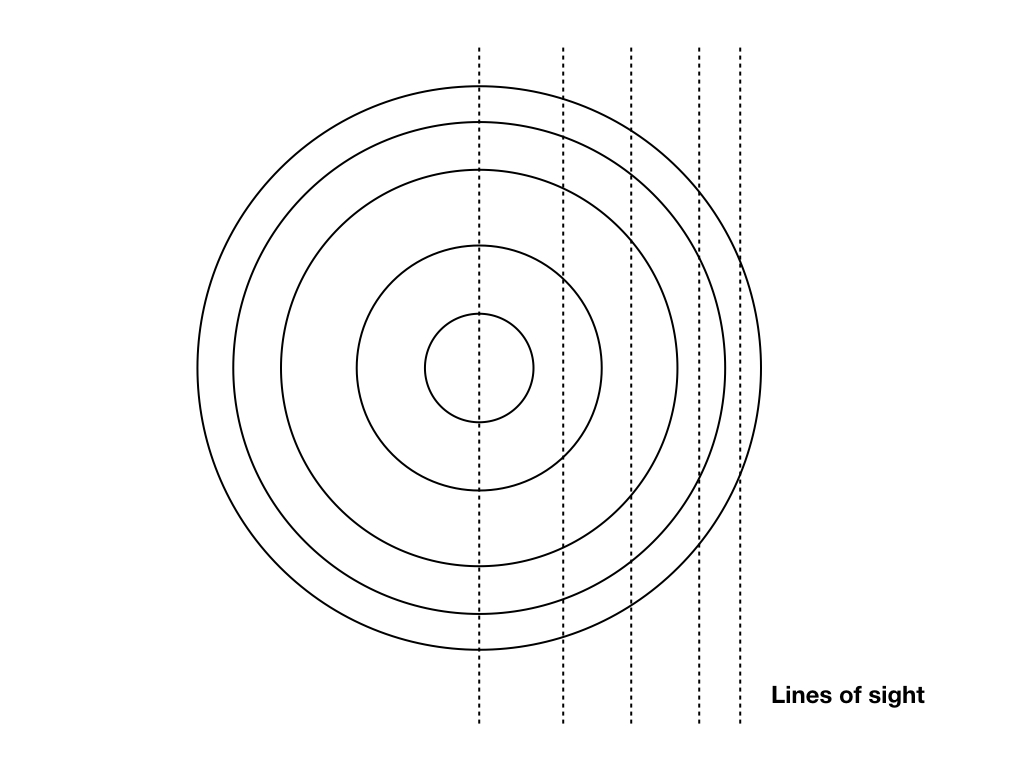} 
\caption{Shell model of the Crab Nebula. We calculate the synchrotron emission along different  lines of sight (dashed lines.)}
 \label{CarbNebulaModel}
\end{figure}

Given the temporal and  the  corresponding spatial evolution of the \Bf, Eq (\ref{Boft}) and the particles' Green's function (\ref{foft}), we  calculate  
the spectral 
 luminosity along a given line of sight  at any moment $t$:
\be
L(\omega, t)=\int_{r_{\min}}^{r_{\max}}  dl \int {N(\gamma, t, r) P(\omega, r,t)} \, d\gamma  
\label{luminosity}
\ee
where the integration path passes through a different shell, see Fig. \ref{CarbNebulaModel}

In practice, we  break the Nebula into a number of thin shells (180 in total in our calculation), and choose shell  spacing equal in observed radii.
The choice of equal spacing in the observed radii is important: equal spacing in presently observed radii corresponds to {\it different} duration of injection time for different shells, see Eq. (\ref{rshell11}). We chose the innermost shell  at 0.100 $R_{PWN,now}$ and each shell has  a width of 0.005 $R_{PWN,now}$.

The ejection time for each shell  is given by Eq.  (\ref{teject}), 
where $B_{now}$ represent the current  magnetic field in the Nebula  and is a free parameter in our model. We then chose 10 lines of sight which  are equally spaced in observed radii, i.e., 0.1 $R_{PWN,now}$, 0.2 $R_{PWN,now}$, ... , 1.0 $R_{PWN,now}$. 
Using $N\left(t\right)=\int G\left(t,t_{inj}\right) dt_{inj}$, for a given injection spectrum,  we know the distribution function at each point in the Nebula at any given time. We can then calculate the spatially resolved synchrotron emissivity (see \S \ref{Step}) and the IC power (see \S \ref{stepIC}).



We adopt the following step-by-step method of fitting the observed spectrum from synchrotron emission: 
\begin{itemize}

\item We estimate $p_{II2}$ from X-ray observations. 

\item We fit the optical index map to estimate $B_{now}$. Stronger $B_{now}$ produces a sharper rise at outer shells and weaker $B_{now}$ produces a milder rise at outer shells. 

\item Once we have the estimate of $B_{now}$, we are able to estimate $\gamma_{I_{max}}$ and $\gamma_{II_{max}}$ according to the broad-band spectrum at the synchrotron limit region, where we expect both components to disappear above 100 MeV. 


\item The requirement  that Component-II does not overshoot  Component-I in the  X-ray region gives a  range of  allowed $\gamma_{II_{break}}$.

\item  We also fit the IR index map of the innermost shell, which is $\alpha \approx 0.3$ for lower frequencies and $\alpha \approx 0.5$ for higher frequencies. This gives $\gamma_{II_{break}}$ and $\gamma_{I_{min}}$.

\item  Given the above estimates,   we are  then able to find the best value of relative normalization factors of Component-I and Component-II.


\end {itemize}

\subsubsection{The IC component}
\label{ic}

Both the particle and the photon distribution within the Nebula are very broad, so that for different parts of the distribution, the IC scattering occurs both in Thomson and Klein-Nishina regimes. The general expression for the differential cross section is \citep[\eg][]{2004vhec.book.....A}
\begin{equation}
\frac{d\sigma_{KN}}{d\Omega} = \frac{3}{16 \pi} \frac{\sigma_{T}}{\left(1+x\left(1-\cos \theta\right)\right)} \left(x\left(1-\cos \theta \right)+\frac{1}{1+x\left(1-\cos \theta \right)}+\cos^2 \theta\right)
\label{dsigma}
\end{equation}
where $x$ is the initial photon energy in units of $m_{e}c^2$, and $\theta$ is the scattering angle in the frame where the electron is initially at rest.

Transformations of the directions and the energies of incoming, scattered photons and the lepton's velocity is a complicated exercise in  Lorentz transformation \citep[\eg][]{1981Ap&SS..79..321A,1996MNRAS.278..525A,1990MNRAS.245..453C}. In particular, \cite {1981Ap&SS..79..321A} derived the angle-averaged scattering rate analytically, and \cite{1990MNRAS.245..453C} re-derived the angle-averaged scattering rate by considering some standard asymptotic forms. In our work, we derived the angle-averaged outgoing photon energy, and  then  calculate  it numerically. 

The notations are the following. In the electron comoving frame $K^{\prime}$,  $x^{\prime}$ is the energy of the  incoming photon, $x_1^{\prime}$ is the energy of the outgoing photon, $\psi^{\prime}$ is the angle between the electron velocity and incoming photon direction, $\psi_1^{\prime}$ is the angle between the electron velocity and outgoing photon direction, $\delta^{\prime}$ is  the azimuthal angle and $\theta^{\prime}$ is  the scattering angle. In the lab frame, we define $x$ as incoming photons energy, $x_1$ as outgoing photons energy, $\psi$ as the angle between the electron velocity and incoming photon direction.

Combining 
Lorentz transformations
\be
 x^{\prime}=\frac{x}{\gamma \left(1+\beta \cos \psi^{\prime}\right)}
\ee
with Compton scattering 
\begin{equation}
 x_{1}^{\prime}=\frac{x^{\prime}}{1+x^{\prime}\left(1-\cos \theta^{\prime}\right)},
\end{equation}
we find 
\begin{equation}
 x_{1}=\frac{x \gamma \left(1+\beta \cos \psi_{1}^{\prime}\right)}{\gamma \left(1+\beta \cos \psi^{\prime}\right) + x \left(1-\cos \theta^{\prime}\right)}
\end{equation}

The geometric relation between scattering angle $\theta^{\prime}$, azimuth angle $\delta^{\prime}$, angle between incoming photon and electron $\psi^{\prime}$ and angle between outgoing photons and electron $\psi_{1}^{\prime}$ is:
\begin{equation}
\cos \psi_{1}^{\prime} = \cos \theta^{\prime} \cos \psi^{\prime} - \sin \theta^{\prime} \cos \delta^{\prime} \sin \psi^{\prime}
\end{equation}
which gives
\begin{equation}
 x_{1}=\frac{x \gamma \left(1+\beta \left(\cos \theta^{\prime} \cos \psi^{\prime} - \sin \theta^{\prime} \cos \delta^{\prime} \sin \psi^{\prime}\right)\right)}{\gamma \left(1+\beta \cos \psi^{\prime}\right) + x \left(1-\cos \theta^{\prime}\right)}
\end{equation}

The Lorentz transformation for angle is $\cos \psi^{\prime}=\frac{\cos \psi-\beta }{1-\beta  \cos \psi }$, thus
\begin{equation}
 x_{1}=\frac{x \gamma \left(1+\beta \left(\cos \theta^{\prime} \frac{\cos \psi-\beta }{1-\beta  \cos \psi } - \sin \theta^{\prime} \cos \delta^{\prime} \sin \psi^{\prime}\right)\right)}{\gamma \left(1+\beta \frac{\cos \psi-\beta }{1-\beta  \cos \psi }\right) + x \left(1-\cos \theta^{\prime}\right)}
\end{equation}

{Then averaging over angle $\delta^{\prime}$ and $\psi$, we have}
\begin{equation}
x_{1}=
\frac{\csc ^2\frac{\theta^{\prime} }{2} \left(\left(\gamma - x \cos\theta^{\prime}\right) \ln \left(\frac{2 \gamma -x \cos \theta^{\prime}+x}{4 \gamma ^2 \left(\frac{1}{2 \gamma }-x \cos\theta^{\prime} +x\right)}\right)+2 x \gamma^2 (1- \cos \theta^{\prime} )\right)}{4 x \gamma}
\label{cos}
\end{equation}
{The Eq. (\ref{cos}) is valid in the limit of $\gamma \gg 1$.}

In order to fit the IC component,  we adopt the step-by-step procedure of fitting the observed spectrum from IC emission:
\begin{itemize}
\label{stepIC}
\item For the sample of Lorentz factor of electrons (say $\gamma$ = 200, $\gamma$ = 400, ...), we calculated the corresponding number density of electrons $n_{e}$, and made a table of value as $n_{e}$ vs. $\gamma$.

\item For the sample of incoming photon energies (say $x=10^{-7}$ eV, $x=2\times10^{-7}$ eV, ...), we calculated the corresponding number of incoming photons $N_{\gamma}$, and made a table of value as $N_{\gamma}$ vs. $x$.

\item For the sample of outgoing photon energies $x_1$, we made a table of $N_{scattered}$ vs. $x_1$, where $N_{scattered}$ is unknown and will be calculated in the following steps.

\item We pick values of $\gamma$, $x$ and $x_1$ from the tables and run the loop (e.g $\gamma=200, x=10^{-7} eV, x_1=10^5 eV$), and we solve Eq. \ref{cos} to find the value of $\cos \theta^{\prime}$.

\item Assuming that the solution of Eq. \ref{cos} is $\cos \theta^{\prime}$ = $S\left(x, x_1, \gamma\right)$, then d$\left(\cos \theta^{\prime}\right) = S_{max}\left(\check{x}, \check{x}_1, \check{\gamma} \right) - S_{min}\left(\check{x}, \check{x}_1, \check{\gamma}\right)$, for $\check{x} \in \left[x-\frac{dx}{2}, x+\frac{dx}{2}\right]$, $\check{x}_1 \in \left[x_1-\frac{dx_1}{2}, x_1+\frac{dx_1}{2}\right]$, $\check{\gamma} \in \left[\gamma-\frac{d\gamma}{2}, \gamma+\frac{d\gamma}{2}\right]$, where $dx$, $dx_1$ and $d\gamma$ are step length in the table.

\item Substitute the value of $x^{\prime}$, $\cos \theta^{\prime}$ and $d\cos \theta^{\prime}$ into Eq. \ref{dsigma}, we can calculate the corresponding differential cross section.

\item Then we substitute the corresponding number density of electrons $n_{e}$ and number particle of incoming photons $N_{\gamma}$ (say the $i^{th}$ row in the table is value $n_{e_{i}}$ and the $j^{th}$ row in the table is value $N_{\gamma_{j}}$),  the collision rate would be $n_{e_{i}} N_{\gamma_{j}} c \sigma$. We need to be aware of that all variable above are in rest frame of electron. So

\item The collision rate in lab frame is $N_{scattered} = n_{e_{i}} N_{\gamma_{j}} c \sigma / \gamma_{i}$.

\item Finally, sum up over the table of value of electrons and multiply the scattered photon frequency, we will find $\left(\nu F\left(\nu\right)\right)_{scattered}$ $\propto$ $\sum_{i,j} \nu x_{1} N_{scattered}$ = $\sum\limits_{i,j} \nu x_{1}  n_{e_{i}} N_{\gamma_{j}} c \sigma / \gamma_{i}$.

\end{itemize}

We verified that the step-by-step procedure described here reproduces a number of analytical results (\eg, IC scattering of mono-energetic seed photons and mono-energetic electrons, IC scattering of mono-energetic seed photons and power law energy distribution electrons).

\subsubsection{The SSC component}
\label{overall}

The model has a number of parameters, \S \ref{param}.
By adopting the step-by-step methods from section \ref{Step} and section \ref{ic}, we calculated the overall spectrum by adding the two synchrotron components and the SSC component. 

The SSC emission is shown as curve 5 in Fig. \ref{starlight}.
  Given that the model is very simple, \eg\ one-dimensional, and spans nearly 20 orders of magnitude in energy and some seven orders of magnitude in flux,  the fits were done "by eye".
 We found the best values of all parameters are $B_{now} = 2.7 \times 10^{-4}$G, $\gamma_{I_{max}} = 3.5 \times 10^9$, $\gamma_{II_{max}} = 8.0 \times 10^9$, $p_{II2} = 2.7$, $\gamma_{I_{min}} = 1.9 \times 10^5$, and $\gamma_{II_{break}} = 2.0 \times 10^6$.
Component-II constitutes about $60\%$ of the ejection energy and Component-I constitutes about $40\%$ of the ejection energy.  The numerical fitting program may be added in further work to improve the precision of parameters value, but for now, our results have good enough precision to demonstrate our model. We summarize all parameters values in Table. \ref{values}. 

\begin{table}[t]
\begin{tabular}{|c|c|c|c|c|c|c|c|c|}
\hline
parameters & $B_{now}$ (G)            & $\gamma_{I_{max}}$ & $\gamma_{II_{max}}$ & $p_{II2}$ & $\gamma_{I_{min}}$ & $\gamma_{II_{break}}$ & $E_{II}/E_{total}$ & $E_{I}/E_{total}$ \\ \hline
values     & $2.7 \times 10^{-4}$ & $3.5 \times 10^9$  & $8.0 \times 10^9$   & 2.7       & $1.9 \times 10^5$ & $2.0 \times 10^6$     & 0.6               & 0.4             \\ \hline
\end{tabular}
\caption{Summary of parameter values. In this table, $B_{now}$ is the magnetic field now. $\gamma_{I_{max}}$ is the maximum Lorentz factor of injected electrons of Component-I. $\gamma_{II_{max}}$ is the maximum Lorentz factor of injected electrons of Component-II. $\gamma_{II_{break}}$ is the middle break Lorentz factor of Component-II, where power law indices are $p_{II1}$ = 1.6 below the $\gamma_{II_{break}}$ and $p_{II2}$ above the $\gamma_{II_{break}}$. $\gamma_{I_{min}}$ is the minimum Lorentz factor of injected electrons of Component-I. $E_{I}$ is the energy of Component-I, $E_{II}$ is the energy of Component-II, and $E_{total}$ is the sum of the energy of Component-I and Component-II.}
\label{values}
\end{table}

\begin{figure}[t]
  \centering
  \includegraphics[width=0.8\textwidth]{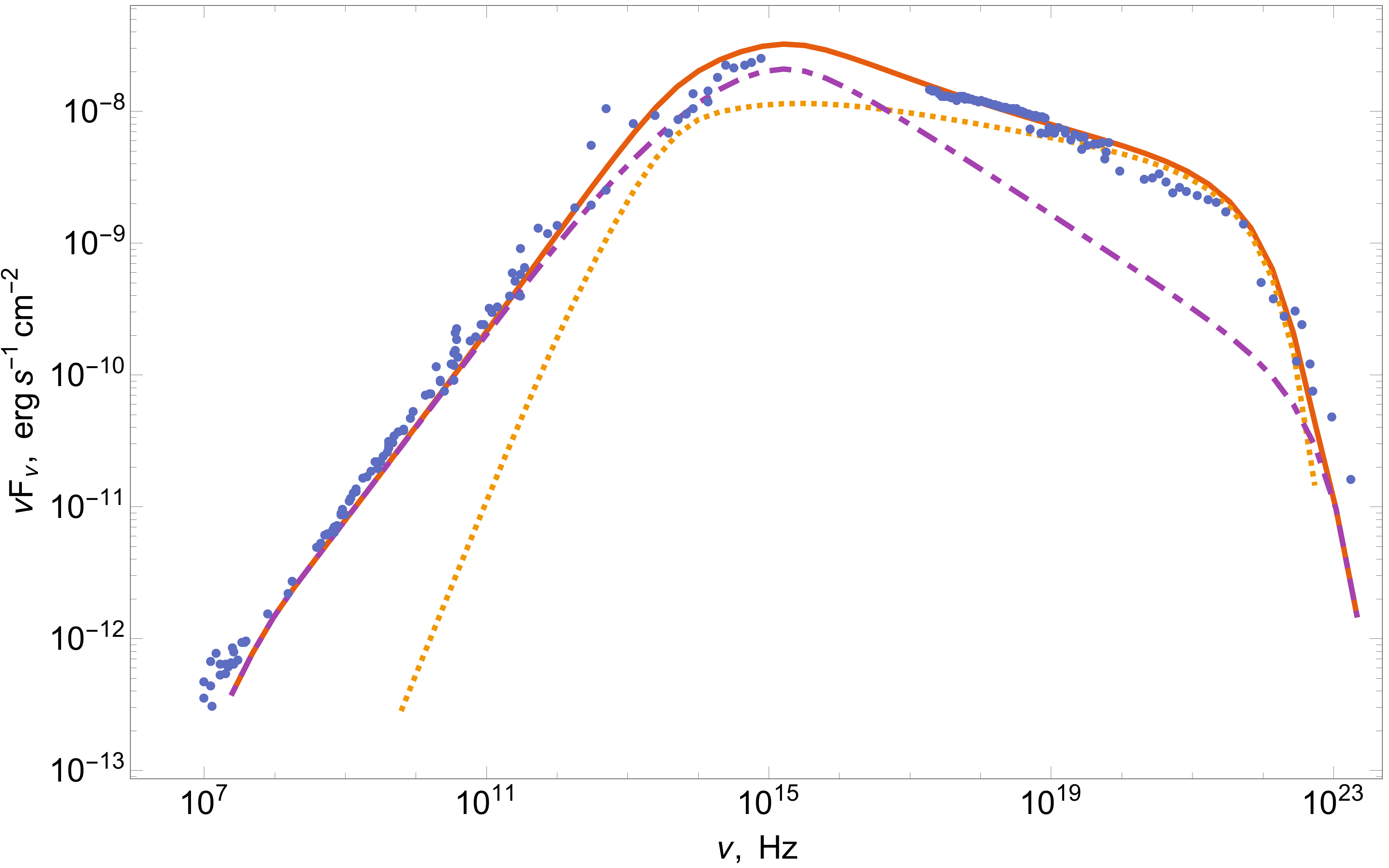}
  \caption{Comparison of observational data \cite{1971IAUS...46...22B,1977A&A....61...99B,2010ApJ...711..417M,1968ApJ...152L..21N,1979PASP...91..436G,2004MNRAS.355.1315G,2006AJ....132.1610T,2001A&A...378..918K} and numerical result for the broad-band spectrum. The dots represent observational data. The red solid line represents the total emission in the model. The purple and yellow dashed line represent Component-I and Component-II.}
  \label{broadband}
 \end{figure}

We then substituted all of parameter values from Table \ref{values} into Eq. (\ref{luminosity}) and calculated the broad-band synchrotron spectrum in Fig. \ref{broadband}, where we present the synchrotron emission from Component-I and Component-II as yellow dotted line and purple dotted-dash line respectively, and their combined contribution as the red solid line. As we can see, the low energy synchrotron emission is dominated by Component-II and high energy synchrotron emission is dominated by Component-I. In the next section \ref{ic}, we will use the broad-band synchrotron spectrum as seed photons for the IC component calculation.

As shown in Fig. \ref{starlight}, our purely SSC emission model with parameter values taken from Table \ref{values} roughly reproduce the current broad-band spectrum. The overall spectrum consists of three parts: Part I: $10^8 - 10^{14}$ Hz is the low energy emission and is dominated by synchrotron emission from Component-II, which has a peak at around $10^{14}$ Hz. $p_{II_{2}}$ does not affect the overall spectrum significantly, however, it will affect IR spectra index map in section \ref{indexmap}. Part II: $10^{16} - 10^{22}$ Hz is the middle energy emission and is dominated by synchrotron emission from Component-I. Part III: $10^{22} - 10^{28}$ Hz is the high energy emission and has a peak around $10^{26}$ Hz. Part III is dominated by SSC emission with taking account synchrotron emission from both Component-I and Component-II as seed photons.

\subsubsection{Dust and starlight contributions}

\begin{figure}[h!]
  \centering
  \includegraphics[width=0.8\textwidth]{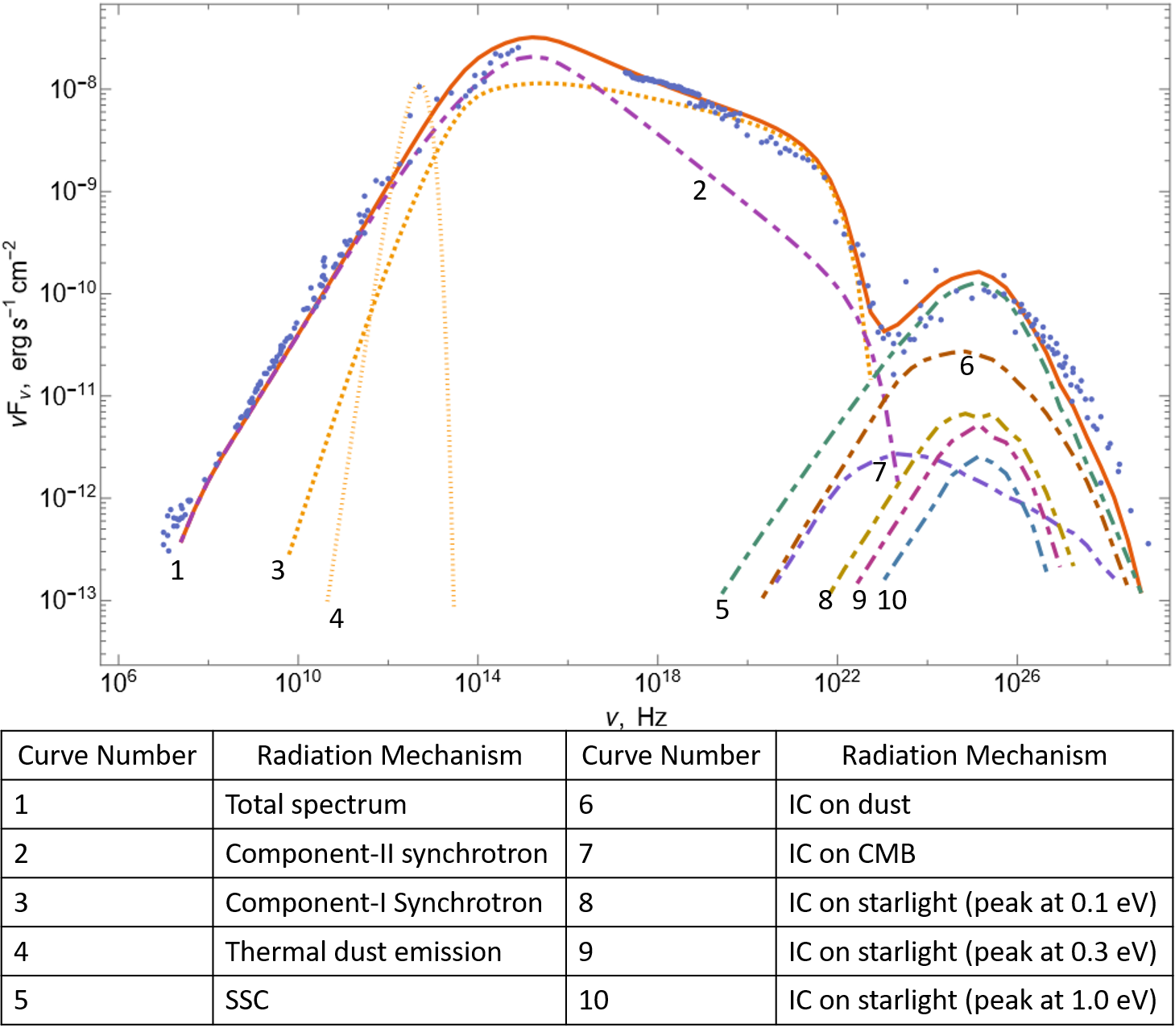}
  \caption{Broad-band spectrum of Crab Nebula. The observational data are showed as blue dots (synchrotron data is same as Fig. \ref{broadband} and we add more data from \cite{2006A&A...457..899A,2008ApJ...674.1037A,2010ApJ...708.1254A} above synchrotron limit). Component-II (curve 2) and Component-I (curve 3) synchrotron emission are taken from Fig. \ref{broadband}. SSC emission is showed as curve 5. IC on thermal dust emission (curve 4) is showed as curve 6. IC on CMB is showed as curve 7. IC on starlight are showed as curve 8 (peak energy at 0.1 eV), curve 9 (peak energy at 0.3 eV) and curve 10 (peak energy at 1.0 eV). The overall total spectrum is showed as curve 1 (here we ignore IC on starlight).}
  \label{starlight}
 \end{figure}

There is a big gap around $10^{23} - 10^{26}$ Hz region between observational data and our numerical SSC emission. In order to fill up this big gap, we consider additional IC photons on CMB and dust. First  we calculated the IC on seed photons, including CMB, Component-I and Component-II. The IC on CMB is showed as curve 7 in Fig. \ref{starlight}. As we can see, additional IC emission on CMB are not able to gives a apparent rise or fill up the gap around $10^{23} - 10^{26}$ Hz region. Thus, we need to add IC emission from dust.

We then consider thermal emission from dust with temperature 62K, and the normalization factor is determined by fitting a small bump in IR band around $5 \times 10^{12}$ Hz. The thermal dust emission is showed as curve 4 in Fig. \ref{starlight}. The associated IC emission gives a comparable contribution and fill up the gap.  See curve 6 in Fig. \ref{starlight} 

Our step-by-step method does not try to fit and calculate two synchrotron components and IC emission at the same time. Fitting-to-all (two synchrotron emission mechanism and IC emission) numerical algorithm with some statistical index checking could be implemented so that we can get better fitting result. But apparently, it cost more time to fit two physical process at the same time. \cite{1968PhRv..167.1159J}, \cite{1970RvMP...42..237B}, \cite{1981Ap&SS..79..321A}, \cite{1990MNRAS.245..453C} proposed different way to calculate IC emission analytically and numerically, however, the way we adopted in this paper is the most acceptable way by trading off time and precision.

Star light photons also have IC emission within nebula, thus we investigate the effect of  IC on star light in this section. We assume that seed photons of IC are from black body  emission (for star light with  different temperatures corresponding to $0.1, \,  0.3$ and $1.0$ eV). Then we adopted our step-by-step method from section \ref{ic} again and calculated the corresponding IC emission.

In Fig. \ref{starlight}, we present IC on star light photons with peak energy at 0.1 eV (curve 8), 0.3 eV (curve 9) and 1.0 eV (curve 10), which are normalized to flux 1.0 eV/cm$^{3}$ at current time. Even for the highest IC emission on  starlight in the case of peak energy at 0.1 eV,  IC on star light are way below the SSC. Thus in later sections, we ignore the IC emission on star light photons.

Finally, the total spectrum is showed as curve 1 in Fig. \ref{starlight} by combining Component-I and Component-II Synchrotron, SSC, IC on thermal dust emission and IC on CMB (here we ignore IC on starlight photons).

\section{Spectral maps in the optical and IR}
\label{indexmap}

The spatial variations of the non-thermal spectrum have been identified as one of the drawbacks of the Kennel \& Coroniti models \citep[][and \S \ref{Introduction}]{2009ApJ...703..662R,2017SSRv..207..175R}: Kennel-Coroniti pure-MHD spherical advection model gives a constant spectral index with a sharp steepening at the edge of the PWN.
Addition of  diffusion on top of Kennel-Coroniti flow  \cite{1972Ap&SS..16...81G,1991ICRC....2..400R,2012ApJ...752...83T,2016MNRAS.460.4135P}, have been proposed to explain the spectral steepening. Yet, the diffusion model cannot predict the change of the source size with photon energy.

Our method has the ability to reproduce the observed spectral index map, which is gradually steepening from the innermost shell to the edge of the PWN. In order to calculate the spectral index map, we consider our shell model in Fig. \ref{CarbNebulaModel}. Each shell has the same parameters but only the injection time is different. The injection time needs be calculated by Eq. (\ref{teject}). For any given injection time, we are able to calculate the emissivity within each shell. By summing up the emission from each shell, we are able to calculate the total emission along each line of sight. 

\begin{figure}[t]
  \centering
  \includegraphics[width=0.8\textwidth]{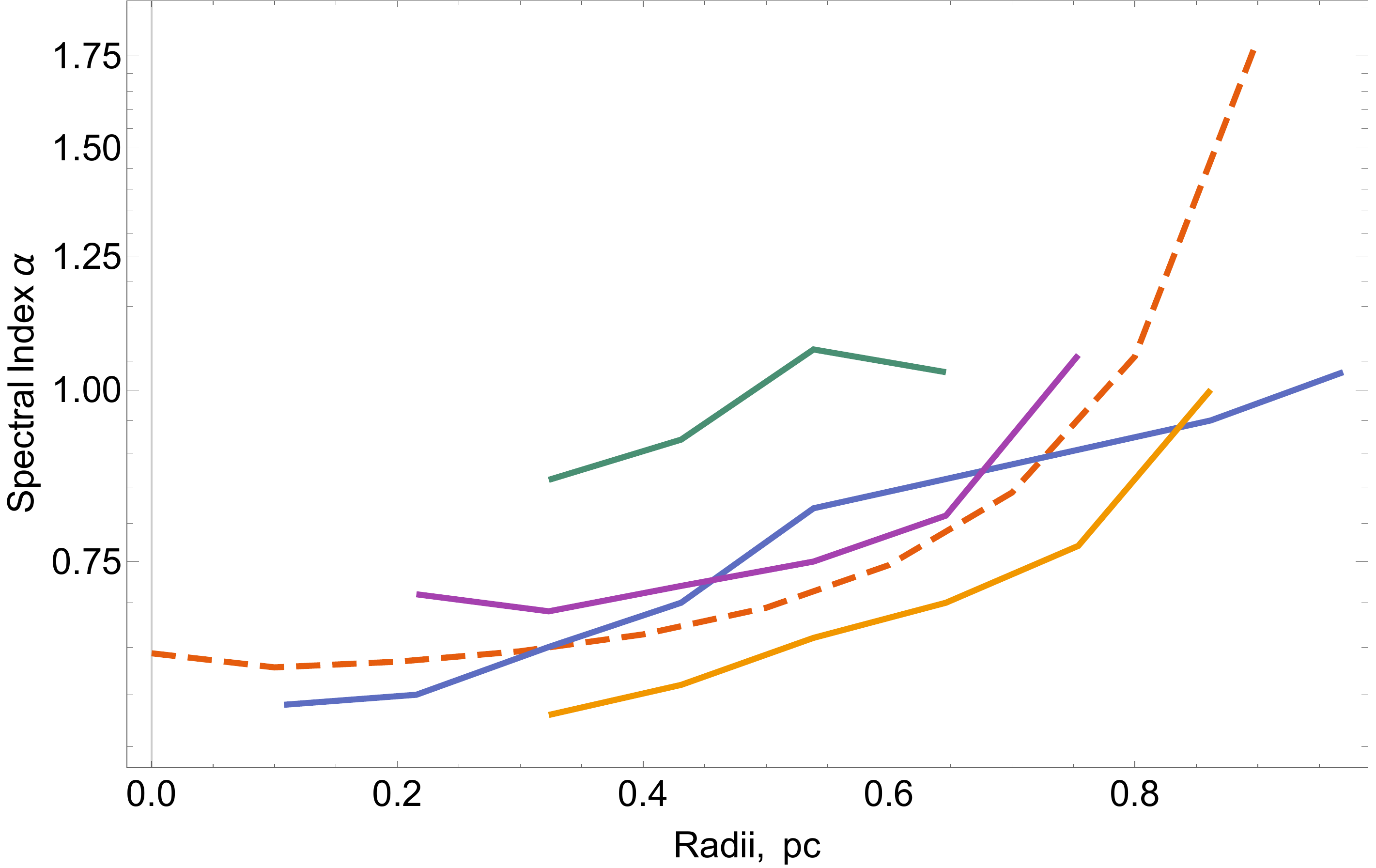}
  \caption{Comparison of observed data \cite{1993A&A...270..370V} and numerical result in the optical region. The wavelength range in the observational data is 0.5364 - 0.9241 $\mu$m. We set the Crab pulsar at 0.0. The green, blue, purple, and orange solid lines represent observational data from west, east, south, and north direction, respectively. The red dashed line represents our numerical result at 0.7 $\mu$m.}
  \label{alpha2}
 \end{figure}

\begin{figure}[t]
  \centering
  \includegraphics[width=0.8\textwidth]{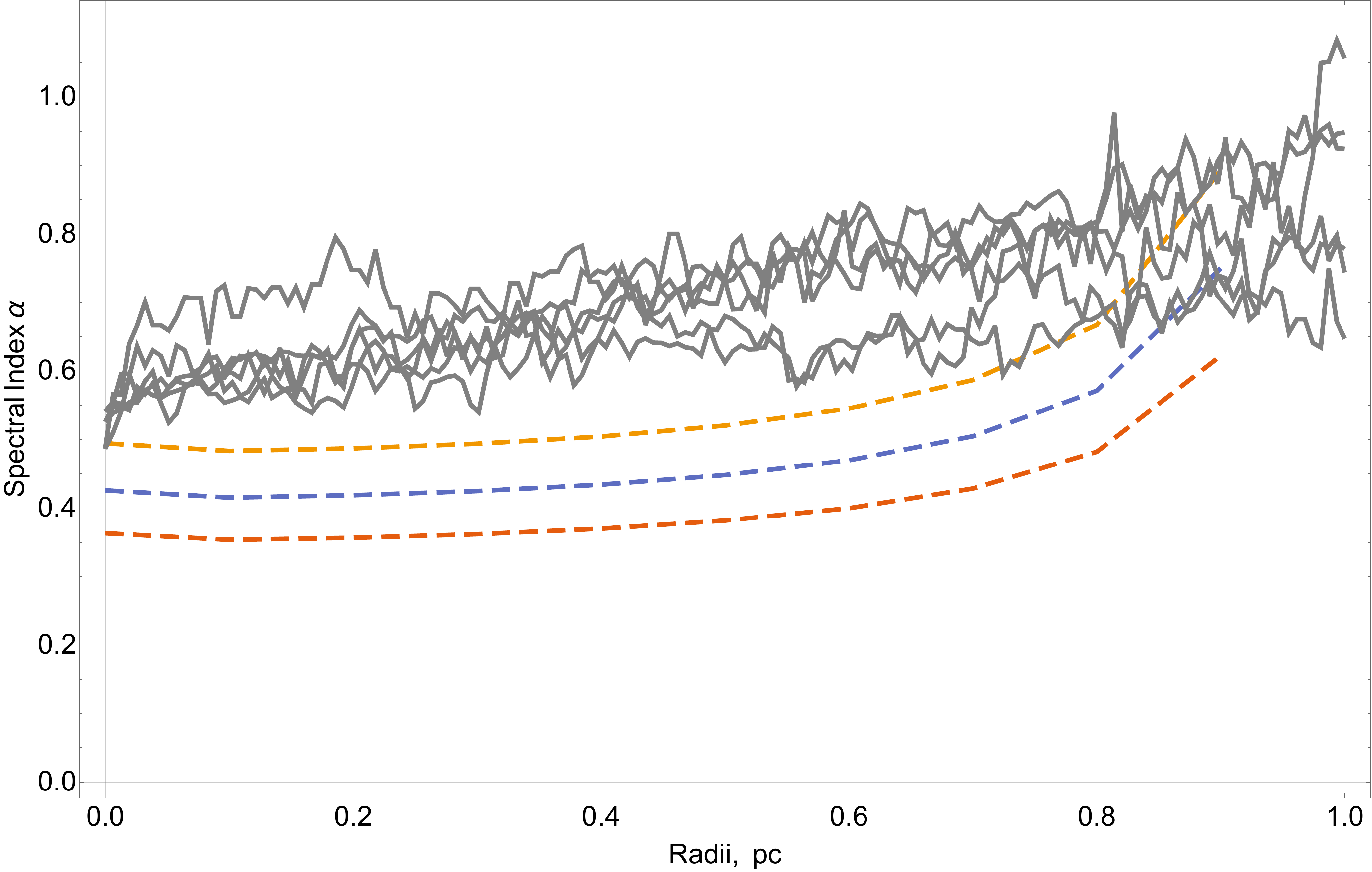}
  \caption{Comparison of the observed data and numerical result in the lower-frequency IR region. The wavelength range in the observational data is 3.6 - 8.0 $\mu$m. We set the Crab pulsar at 0.0. The solid lines represent observational data along different directions. The red dashed lines represents our numerical result at 7.9 $\mu$m. The blue dashed line represents our numerical data at 5.3 $\mu$m. The orange dashed line represents our numerical data at 3.5 $\mu$m. Even though we are trying to match the innermost shell index instead of the whole index map, the trend seen in the whole index map is similar to our numerical model.}
  \label{ir2}
 \end{figure}

\begin{figure}[t]
  \centering
  \includegraphics[width=0.8\textwidth]{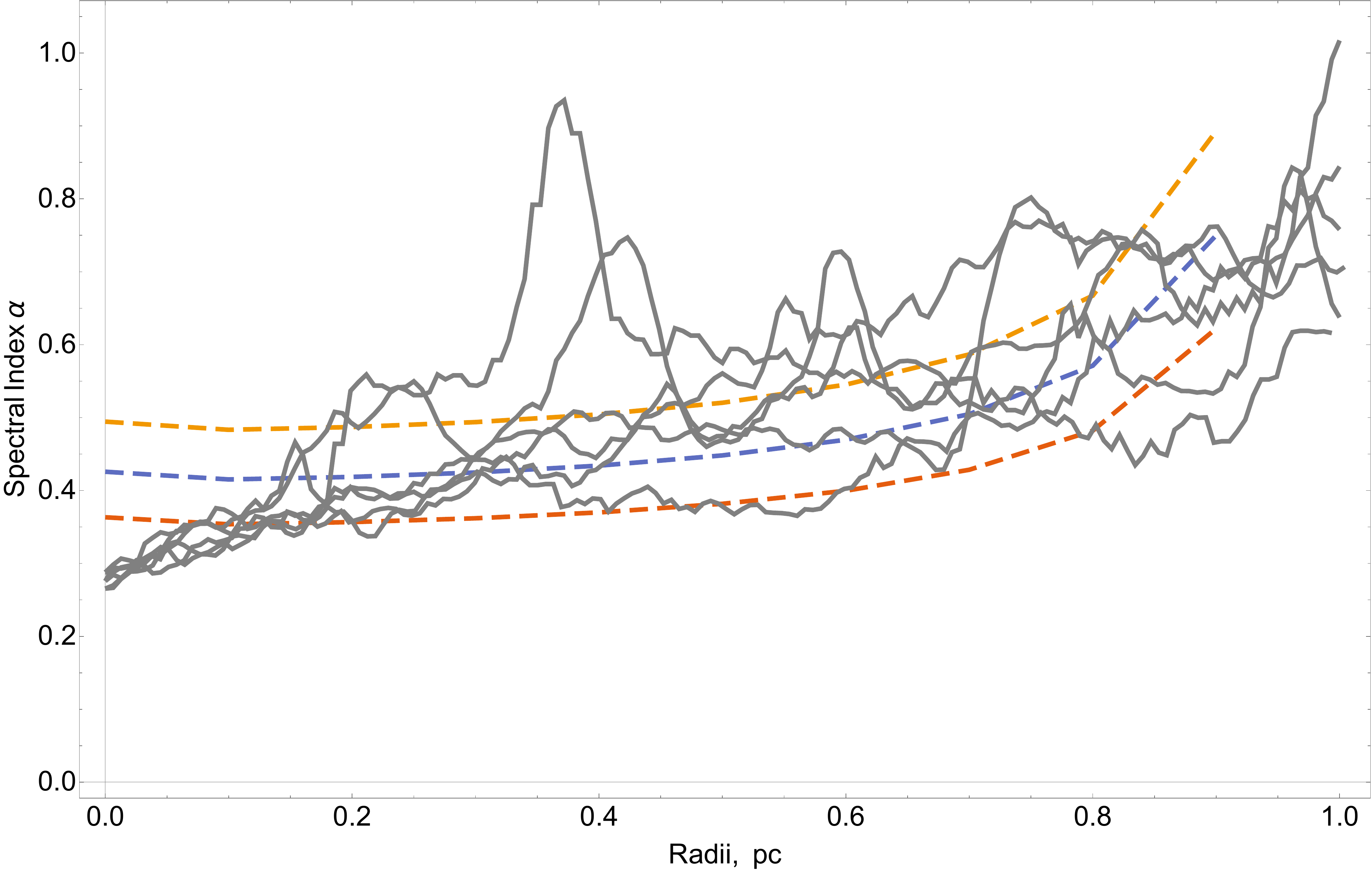}
  \caption{Comparison of observed data and numerical result in the higher frequency IR region. The wavelength range in the observational data is 3.6 - 4.5 $\mu$m. We set the Crab pulsar at 0.0. The solid lines represent observational data along different directions. The red dashed lines represents our numerical result at 7.9 $\mu$m. The blue dashed line represents our numerical data at 5.3 $\mu$m. The orange dashed line represents our numerical data at 3.5 $\mu$m. Even though we are trying to match the innermost shell index instead of the whole index map, the trend seen in the whole index map is similar to our numerical  model.}
  \label{ir1}
 \end{figure}

In our work, we calculate the emission along each line of sight in the IR (7.9 $\mu$m, 5.3$\mu$m, and 3.5$\mu$m) and optical wavelengths (0.7$\mu$m), and then we use them to plot the spectral index map at each frequency. Results are presented in   Figs. \ref{alpha2}--\ref{ir1}.

 Figs. \ref{alpha2}--\ref{ir1} show that the spectral index maps from radio to IR are consistent with observational results, thus demonstrating that our model can generally reproduce the evolution of the spectral indices in IR and optical.


\section{Conclusion}

In this paper, following \cite{2019MNRAS.489.2403L},  we further develop  a turbulent model of the Crab Nebula, and by extension, of PWNe in general. We demonstrate  that developed turbulence in the magnetized post-shock wind can consistently resolve a number of problems of the Kennel and Coroniti model, both theoretical and observational. Turbulence and ensuring reconnection destroys  the magnetic flux, resolving the long-standing sigma-paradox, explains the origin  and spectrum of radio electrons, gamma-ray flares, and the spectral evolution of the flow. With a simple 1D model, we are able to fit, within a factor of few, the   broadband  spectrum that stretches over 20 orders of magnitude in frequency. 
Importantly, the model suggests that reconnection is an important   particle acceleration mechanism in a major astrophysical object - and, by  extension, may be important/dominant in other astrophysical high-energy sources.

  We advocate two acceleration mechanisms that produce two separate particle components:  
  Component-I originates from particles accelerated at the   terminate shock, presumably via the Fermi-I acceleration mechanism.   Component-I dominates from optical to X-ray wavelengths and produces mostly the bright X-ray torus.  Component-II is generated by magnetized turbulence that produces  reconnecting current sheets of different sizes in the bulk of the Nebula. Particles are then accelerated by magnetic reconnection in the current layers and by scattering off turbulent fluctuations. 
Both the  hard  radio spectrum of Component-II and the  requirement that rare reconnection events produce gamma-ray flares,  requires  regions with high
magnetization, $\sigma \gg 1$. 
  
  Thus, we argue that the radio emitting leptons are accelerated by the same mechanism as  GeV emitting leptons, but are different from the X-ray emitting ones. This is different from  \cite{2014MNRAS.438.1518O,2015MNRAS.449.3149O} where the two populations were non-overlapping in energy. 
  One of the major advantages of our model is that it is physically motivated, and not just an {\it ad hoc} parametrization.
  
  The model also explains low injection \Lf\ for the Component-II, $\gamma_{II, min}$ \cite[see more detailed discussion in][]{2019MNRAS.489.2403L}. At mid-latitudes the pulsar wind is relatively slow, $\gamma _w\sim 10^2$, and highly magnetized,  $\sigma _{w} \sim 10^3$.  Thus, the total energy per particle (in terms of $m_e c^2$) is $\gamma_p \sim \gamma_w \sigma_w\sim 10^5$. Within the striped part of the wind this total energy is given to the particles, producing the break at $\gamma_{I, min}$. At the intermediate attitudes, where the wind is not striped, only the bulk energy is thermalized, giving  $\gamma_{II,min} \sim  \gamma_w \sim 10^2$.

There is a number of issues that remain to be resolved. First, our 1D model naturally cannot reproduce azimuthal variations in the properties of the Crab Nebula. Presumably they originate due to  intrinsic anisotropy of the wind and mildly relativistic velocities  (and corresponding Doppler corrections) of the shocked flow in the innermost parts of the Nebula.

{A more accurate evaluation of the particle energization near the cut-off energy would require a kinetic equation that also includes the effect of particle diffusion. In future work, we want to develop a more refined kinetic model that includes particle diffusion. Synchrotron radiation losses could also be added in Eq. (32). However, the synchrotron cooling of the radio electrons is negligible in the Crab nebula. Particle acceleration by reconnection electric fields also do not suffer significant synchrotron losses since the particle pitch angle is aligned to the magnetic field. On the other hand, the synchrotron losses in Fermi II acceleration would become significant at much higher particle energies. We intend to explore their role with particle-in-cell simulations in the next works.}

The main theoretical  unsolved problem, that the current model depends on is the suggestion that magnetic reconnection can indeed produce a spectrum with $p=1.6$, \S\ref{reconnectingturbulence}. 
Another issue is the shear number of radio emitting electrons \citep{1999A&A...346L..49A}.

LC acknowledges support from DoE DE-SC0016542, NSF ACI-1657507, and NASA ATP NNX17AG21G. ML acknowledges support by 
NASA grant 80NSSC17K0757 and  NSF grants 10001562 and 10001521. ML and YL acknowledge support by Purdue Research Foundation. We would like to thank Steve Reynolds for comments.

\bibliographystyle{apj}
  \bibliography{all,/Users/maxim/Home/Research/BibTex,turb_bib}

\appendix
\end{document}